# Breathers and 'black' rogue waves of coupled nonlinear Schrödinger equations with dispersion and nonlinearity of opposite signs

Jin Hua LI *, Hiu Ning CHAN †,

Kin Seng CHIANG ‡, Kwok Wing CHOW †§

* = School of Physics and Optoelectronic Engineering, Nanjing University of Information Science and Technology, Nanjing 210044, China

† = Department of Mechanical Engineering, University of Hong Kong, Pokfulam, Hong Kong

‡ = Department of Electronic Engineering, City University of Hong Kong, Kowloon, Hong Kong

§ = Corresponding author

Email: kwchow@hku.hk    Fax: (852) 2858 5415

# ABSTRACT

Breathers and rogue waves of special coupled nonlinear Schrödinger systems (the Manakov equations) are studied analytically. These systems model the orthogonal polarization modes in an optical fiber with randomly varying birefringence. Unlike the situation in a waveguide with zero birefringence, rogue waves can occur in these Manakov systems with dispersion and nonlinearity of opposite signs, provided that a group velocity mismatch is present. The criterion for the existence of rogue waves correlates exactly with the onset of modulation instability. Theoretically the Hirota bilinear transform is employed and rogue waves are obtained as a long wave limit of breathers. In terms of wave profiles, a 'black' rogue wave (intensity dropping to zero) and the transition to a four-petal configuration are identified. Sufficiently strong modulation instabilities of the background may overwhelm or mask the development of the rogue waves, and such thresholds are correlated to actual physical properties of optical fibers. Numerical simulations on the evolution of breathers are performed to verify the prediction of the analytical formulations.

## 1. Introduction

Rogue waves are large amplitude displacements which suddenly appear in an otherwise calm situation. They are localized in both time and space, and are frequently characterized as 'extreme or rare events in physical systems'. These waves were first recognized nearly a century ago in the maritime community in the settings of oceanic waves [1], and obviously pose great danger to shipping and offshore structures. After the recent detection of optical rogue waves in a microstructured fiber [2], these unexpectedly large displacements from a tranquil background have commanded increasing attention in optics and other fields of physics [3, 4].

The nonlinear Schrödinger equation (NLSE) is a commonly used model in the analysis of nonlinear wave propagation. In the optical fiber setting, the dynamics of a slowly varying envelope of the electric field is governed by the competing influence between second order dispersion and Kerr (cubic) nonlinearity [5]. Rogue wave modes then only occur in the anomalous dispersion regime, i.e. nonlinearity and dispersion of the same sign, where modulation instability (MI) is present [6]. MI represents an exponential growth of small disturbances on a plane wave background arising from the interplay between dispersion and nonlinearity, and results in a broadband output spectrum from an input of a relatively narrow range of frequencies. In the present optical context, anomalous (normal) dispersion regime corresponds to the nonlinear Schrödinger equation with nonlinearity and dispersion of the same (opposite) signs respectively.

MI is closely related to the dynamics of a breather, a pulsating mode of wave propagation. For the NLSE relevant to a temporal soliton in a fiber [5], two classes of breathers are possible: the Kuznetsov-Ma breather, which is periodic in space (the propagation variable) but localized in time, and the Akhmediev breather, which is periodic in time but localized in space [7, 8]. The long wave limits of both breathers yield the Peregrine soliton, the widely used algebraic model of a rogue wave [9, 10].

For coupled NLSE systems [11], MI can occur even in the normal dispersion regime, where dispersion and nonlinearity are of opposite signs [12–17]. This may happen if the cross phase modulation (XPM) coefficient is larger than that due to self phase modulation (SPM). Alternatively the presence of a group velocity mismatch might produce the same effect. However, analysis on breathers and rogue waves in such regimes has not received much attention, even though studies of rogue waves in the anomalous dispersion regime have been making steady progress [18–20]. Recent works highlighted the existence of rogue waves for coupled NLSE in the normal dispersion regime [21, 22], but apparently no investigation on breathers under such circumstances has ever been taken.

The goal of the present work is to examine theoretically the dynamics of breathers and rogue waves, using a coupled NLSE system with identical SPM and XPM coefficients, i.e. the Manakov model. The relation among MI, breathers, and rogue waves will be clarified. The physics revealed can be realistically interpreted in terms of the propagation of the orthogonal polarization modes in an optical fiber with birefringence [23–25], as rogue waves have been observed in fibers. Furthermore, the line of investigation developed here can be readily extended to other nonlinear evolution systems, e.g. nonlinear Schrödinger system with external potential [26] or quintic nonlinearity [27].

The plan of the paper can now be described. The general formulation of a breather for the Manakov model in the normal dispersion regime is first described, and in particular an Akhmediev breather is calculated analytically (Section 2). A rogue wave is then obtained as a limiting case (Section 3). Conditions of MI for small perturbation wavenumber correlate exactly with the existence criterion of rogue waves (Section 4). Numerical simulations of stabilities confirmed the predictions of these analytical calculations (Section 5). Conclusions are drawn (Section 6).

## 2. Breathers for the Manakov model in the normal dispersion regime

The Manakov model can be obtained from the coupled NLSEs for the envelopes of two orthogonal polarization modes of an optical fiber [23–25], where rapidly varying four-wave-mixing (coherent coupling) terms are ignored and the birefringence in the fiber is assumed to vary randomly along the fiber. Rotation of the Stokes vectors and averaging over the Poincaré sphere lead to

$$i\frac{\partial A}{\partial z} - \frac{\partial^2 A}{\partial t^2} + \sigma(|A|^2 + |B|^2)A = 0, \quad i\frac{\partial B}{\partial z} - \frac{\partial^2 B}{\partial t^2} + \sigma(|A|^2 + |B|^2)B = 0, \quad (1)$$

where $A$, $B$ are the slowly varying envelopes of the two orthogonal polarization modes, $z$, $t$, $\sigma$ are the propagation distance, retarded time and strength of the cubic nonlinearity respectively.

Although the Darboux transformation is used widely in the literature, we shall employ the Hirota bilinear transform here [28, 29]. The bilinear method has been proven effective in finding solitons for over forty years and most integrable evolution equations possess bilinear forms:

$$A = \frac{\rho \exp(2i\sigma\rho^2 z)g}{f}, \quad B = \frac{\rho \exp(ist + is^2 z + 2i\sigma\rho^2 z)h}{f}, \quad g, h \text{ complex}, f \text{ real}, \quad (2)$$

$$(iD_z - D_t^2)g \cdot f = 0, \quad (iD_z - D_t^2 - 2siD_t)h \cdot f = 0, \quad (D_t^2 - 2\sigma\rho^2)f \cdot f = -\sigma\rho^2(|g|^2 + |h|^2) \quad (3)$$

where $D$ denotes the Hirota operator defined by

$$D_t^m D_z^n F \cdot G = \left(\frac{\partial}{\partial t} - \frac{\partial}{\partial t'}\right)^m \left(\frac{\partial}{\partial z} - \frac{\partial}{\partial z'}\right)^n F(z,t)G(z',t')\bigg|_{z'=z, t'=t}.$$

The parameter $s$ represents a mismatch in the group velocity between the two polarization modes, as a simple transformation $A = u$, $B = \exp(ist + is^2 z)v$ can recast Eq. (1) as

$$iu_z - u_{tt} + (uu^* + vv^*)u = 0, \quad iv_z - 2isv_t - v_{tt} + (uu^* + vv^*)v = 0,$$

where $s$ is readily recognized as a group velocity mismatch parameter. The quantity $\rho$ denotes the amplitude of the background continuous wave, assumed to be identical for both polarizations for simplicity. Situations with distinct background amplitudes will be left for future studies.

The breather can now be obtained by considering a 2-soliton solution where the frequencies and wave numbers of the solitons are complex conjugate pairs:

$$\begin{aligned}
f &= 1 + \exp(\omega t - Kz + \zeta^{(1)}) + \exp(\omega^* t - K^* z + \zeta^{(2)}) \\
&\quad + M \exp[(\omega + \omega^*)t - (K + K^*)z + \zeta^{(1)} + \zeta^{(2)}]
\end{aligned}$$

$$\begin{aligned}
g &= 1 + \gamma_1 \exp(\omega t - Kz + \zeta^{(1)}) + \gamma_2 \exp(\omega^* t - K^* z + \zeta^{(2)}) \\
&\quad + M\gamma_1\gamma_2 \exp[(\omega + \omega^*)t - (K + K^*)z + \zeta^{(1)} + \zeta^{(2)}]
\end{aligned} \qquad (4)$$

$$\begin{aligned}
h &= 1 + \delta_1 \exp(\omega t - Kz + \zeta^{(1)}) + \delta_2 \exp(\omega^* t - K^* z + \zeta^{(2)}) \\
&\quad + M\delta_1\delta_2 \exp[(\omega + \omega^*)t - (K + K^*)z + \zeta^{(1)} + \zeta^{(2)}]
\end{aligned}$$

with ω, $K$ being complex frequency and wavenumber respectively, $\zeta^{(1)}$, $\zeta^{(2)}$ being phase factors and

$$\gamma_1 = \frac{iK - \omega^2}{iK + \omega^2}, \gamma_2 = \frac{iK^* - (\omega^*)^2}{iK^* + (\omega^*)^2}, \delta_1 = \frac{iK + 2si\omega - \omega^2}{iK + 2si\omega + \omega^2}, \delta_2 = \frac{iK^* + 2si\omega^* - (\omega^*)^2}{iK^* + 2si\omega^* + (\omega^*)^2},$$

$$M = \frac{(\omega K^* - \omega^* K)^2 + (\omega\omega^*)^2(\omega - \omega^*)^2}{(\omega K^* - \omega^* K)^2 + (\omega\omega^*)^2(\omega + \omega^*)^2}.$$

The dispersion relation is

$$4\sigma\rho^2\omega^2(K^2 + \omega^4 + 2s\omega K + 2s^2\omega^2) = (K^2 + \omega^4)\{(K + 2s\omega)^2 + \omega^4\}. \tag{5}$$

An Akhmediev breather in the present context will be strictly periodic in time and localized in space [7, 8, 30], the dispersion relation will simplify to

$$\omega = i\omega_0 \text{ (}\omega_0 \text{ real)}, \qquad K = \omega_0\left(\pm\sqrt{2\mu_2 - \mu_1} - is\right), \tag{6}$$

$$\mu_1 = \omega_0^2 + s^2 + 2\sigma\rho^2, \quad \mu_2 = \sqrt{s^2\omega_0^2 + 2s^2\sigma\rho^2 + (\sigma\rho^2)^2}. \tag{7}$$

For a non-singular wave profile, $K$ must have a nonzero real part. Eq. (6) implies a condition on the group velocity mismatch,

$$0 < s^2 - \omega_0^2 < 4\sigma\rho^2, \tag{8}$$

as the criterion for the existence of breathers.

## 3. Rogue waves

*3.1 Analytical formulation*

A rogue wave will arise when the period of an Akhmediev breather becomes indefinitely large. Theoretically, a zero frequency limit is taken ($\omega_0 \to 0$ in Eqs. (4 – 8)):

$$K = \omega_0 K_0 + O(\omega_0^3), \quad \text{Re}(K_0) = a, \quad \text{Im}(K_0) = b, \tag{9}$$

$$a = \pm\sqrt{-2\sigma\rho^2 - s^2 + 2\sqrt{(\sigma\rho^2)^2 + 2s^2\sigma\rho^2}}, \quad b = -s. \tag{10}$$

The criterion for the existence of rogue wave follows from Eq. (8) as

$$s^2 < 4\sigma\rho^2. \tag{11}$$

To derive the rogue wave mode analytically, one takes the phase factors $\exp(\zeta^{(1)}) = \exp(\zeta^{(2)}) = -1$ in Eq. (4) and expands $f$, $g$, and $h$ for small $\omega_0$, following analogous procedure in deriving rogue waves for derivative NLSE [31] and long wave-short wave resonance model [32]:

$$A = \rho \exp(2i\sigma\rho^2 z)\left\{1 + \frac{4a^2[-1 + (-st - s^2 z + a^2 z)i]}{(a^2 + s^2)[a^2(t + sz)^2 + a^4 z^2 + 1]}\right\}, \tag{12}$$

$$B = \rho \exp(ist + is^2 z + 2i\sigma\rho^2 z)\left\{1 + \frac{4a^2[-1 + (st + a^2 z + s^2 z)i]}{(a^2 + s^2)[a^2(t + sz)^2 + a^4 z^2 + 1]}\right\}. \tag{13}$$

Similar to the Peregrine soliton of the single component NLSE, this rogue wave is also localized algebraically in both space and time. However, there are striking differences. At the center of the rogue wave ($t = z = 0$ here), the amplitude is less than $\rho$, and hence this mode is a 'dark' rogue wave. Theoretically, a real positive value of the parameter $a$ is necessary (otherwise Eqs. (12, 13) are just plane waves for $a = 0$), and the necessary condition is then Eq. (11).

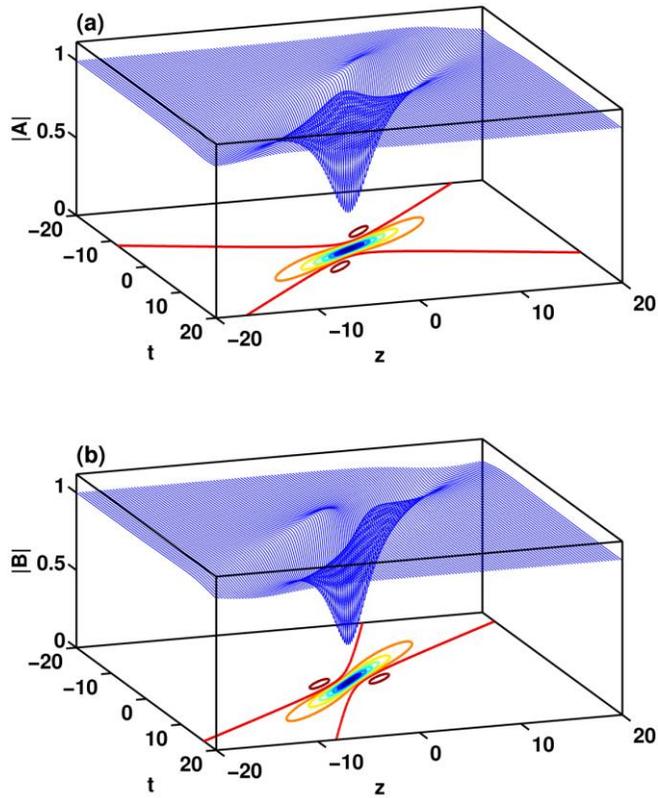

FIG. 1. (Color online) A dark rogue wave pair for the Manakov model in the normal dispersion regime with $\rho = 1$, $s = 1$, and $\sigma = 0.5$ for the two components (a) $|A|$ and (b) $|B|$. There is just one single depression at the center.

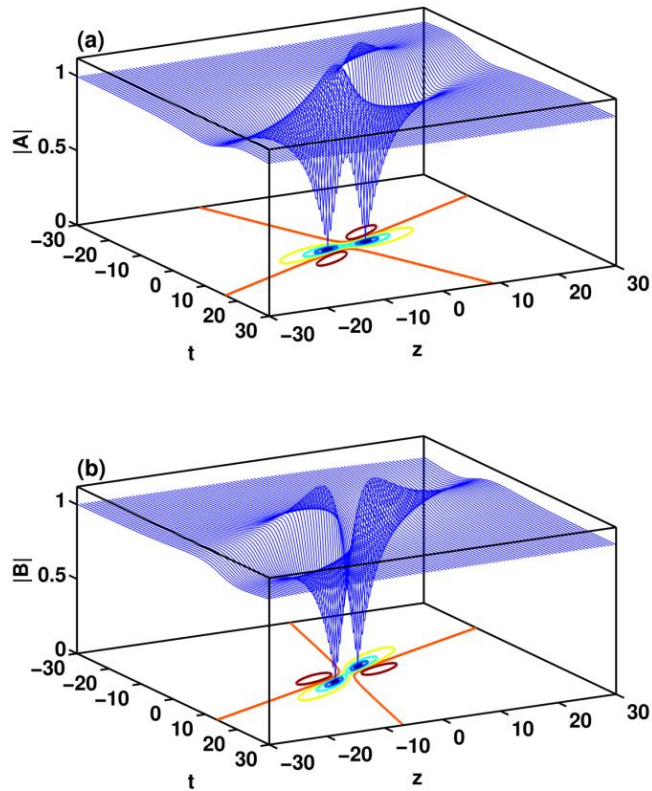

FIG. 2. (Color online) A rogue wave pair in a four-petal configuration for the Manakov model in the normal dispersion regime with $\rho = 1$, $s = 0.5$ and $\sigma = 0.5$ for the two components (a) $|A|$ and (b) $|B|$.

*3.2 Wave profiles*

For NLSE of a single component in the anomalous dispersion regime, the rogue wave consists of a single 'elevation' accompanied by two 'depressions', with the peak of the elevation being three times the background. For two coupled NLSEs in the normal dispersion regime, the main displacements for the present case of equal background consist of depressions below the mean level in the center (a dark rogue wave, Fig. 1). Furthermore, profiles of both components must be of a similar nature, and thus combinations of 'dark-bright' rogue waves are not permitted.

To prove these assertions, first we recognize that $t = z = 0$ is a turning point of the function

$$L = \frac{|A|^2}{\rho^2}.$$

The second derivatives at the stationary point $(t, z) = (0, 0)$ are given by

$$L_{xx}\big|_{(0,0)} = -\frac{48a^4(a^2 - s^2)}{(a^2 + s^2)^2}, \quad (L_{xx}L_{tt} - L_{xt}^2)\big|_{(0,0)} = \frac{256a^{10}(a^2 - 3s^2)(3a^2 - s^2)}{(a^2 + s^2)^4}.$$

The second derivatives at the stationary point $(t, z) = (0, 0)$ for $B$ are the same and hence the profiles must be of a similar nature.

The precise profiles are determined by the signs of these second derivatives. For a local maximum or a minimum, $(a^2 - 3s^2)(3a^2 - s^2)$ must be positive. As $s^2 < a^2/3$ is inadmissible due to Eq. (10), the range of group velocity difference for rogue wave modes to occur is $s^2 > 3a^2$, and the turning point is a minimum ($L_{xx} > 0$), i.e. a dark rogue wave. Since $\int |A|^2 dt$ and $\int |B|^2 dt$ are conserved quantities of this dynamical system, two maximum points emerge above the mean level in the vicinity to compensate for the drop in the intensity at (0, 0). Physically as $s^2$ goes from infinity to $3a^2$, the depression at (0, 0) deepens until the intensity reaches zero at $s^2 = 3a^2$, i.e. a 'black' rogue wave.

For $3a^2 > s^2 > a^2$, theoretically the point (0, 0) transforms from a minimum to a saddle point. Physically the minimum splits into two smaller units, and these units move away from each other in opposite directions, and the wave profile now displays a 'four-petal' pattern (Fig. 2). Regarding the 'elevation' portions of the wave profile, these peaks approach the value $\sqrt{2}\rho$ as $a^2$ tends to $s^2$, in sharp contrast with the Peregrine breather which has a maximum of $3\rho$.

## 4. Modulation instability

The connections between MI and rogue waves have been pursued in the literature, e.g. single component NLSE [33], coupled NLSEs (Manakov model) in the normal dispersion [22] and derivative NLSE [31]. Here we extend the consideration to breathers too. First a quick outline of MI of Eq. (1) is presented. Normal modes for small perturbations of continuous waves of Eq. (1) are sought:

$$A = \rho \exp(2i\sigma\rho^2 z)(1 + \varphi), \quad B = \rho \exp(ist + is^2 z + 2i\sigma\rho^2 z)(1 + \psi), \tag{14}$$

$$\varphi = F_1\exp[i(rt-kz)] + G_1\exp[-i(rt-kz)], \quad \psi = F_2\exp[i(rt-kz)] + G_2\exp[-i(rt-kz)]. \tag{15}$$

Nontrival solutions of $F_1$, $G_1$, $F_2$, and $G_2$ lead to the dispersion relation

$$k = \begin{cases} -sr \pm r\sqrt{r^2 + s^2 + 2\sigma\rho^2 + 2\sqrt{r^2 s^2 + 2s^2 \sigma\rho^2 + \sigma^2 \rho^4}} \\ -sr \pm r\sqrt{r^2 + s^2 + 2\sigma\rho^2 - 2\sqrt{r^2 s^2 + 2s^2 \sigma\rho^2 + \sigma^2 \rho^4}} \end{cases}.$$

MI will occur for complex values of $k$, i.e. values of $r$ which fall in the range of

$$s^2 - 4\sigma\rho^2 < r^2 < s^2. \tag{16}$$

The gain is then given by

$$\text{Gain} = |\text{Im}(k)|. \tag{17}$$

The MI gain spectrum generally increases with the coefficient of cubic nonlinearity ($\sigma$), the group velocity mismatch parameter ($s$), and the amplitude of the background continuous wave ($\rho$) (Fig. 3).

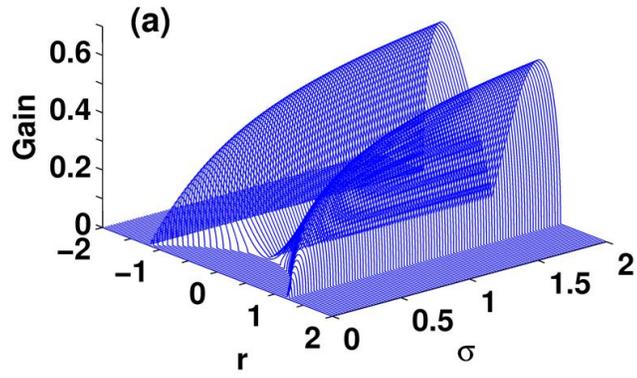

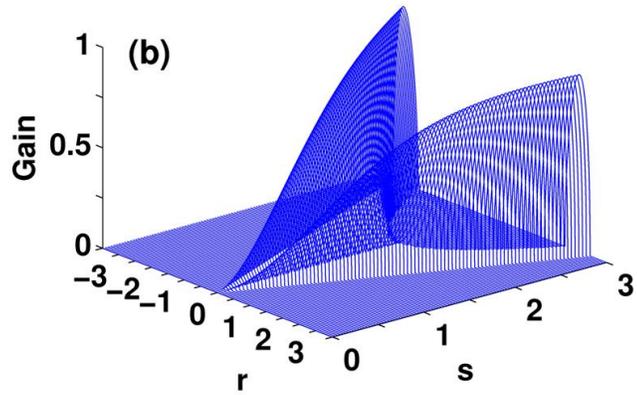

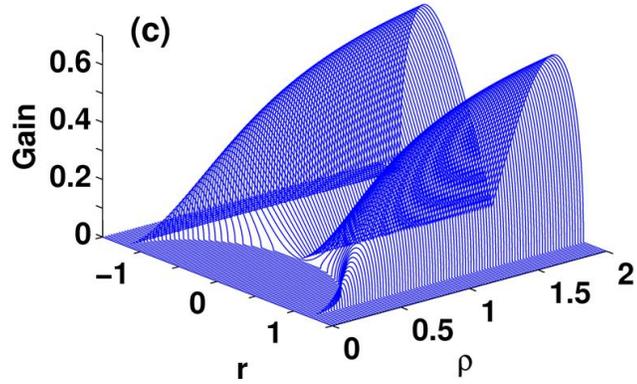

FIG. 3. (Color online) The variations of MI gain spectrum with (a) $\sigma$ for $s = 1.2$, $\rho = 1$; (b) $s$ for $\sigma = 1$, $\rho = 1$; and (c) $\rho$ for $\sigma = 1$, $s = 1.2$.

When $s^2 < 4\sigma\rho^2$ holds, any positive perturbation frequency $r$ in the range of $0 < r < s$ will satisfy Eq. (16), and the zero frequency limit ($r \to 0$) falls within the range of MI. However, in the opposite case, when $s^2 > 4\sigma\rho^2$ holds, the range of unstable perturbation frequency $r$ is restricted to $s^2 - 4\sigma\rho^2 < r^2$, i.e. bounded away from zero. Hence a long wave limit ($r \to 0$) to derive a rogue wave will go beyond the regime of MI. Consequently, the lack of rogue wave modes for the parameter regime $s^2 - 4\sigma\rho^2 > 0$ is not surprising.

This analysis now establishes the connection between the existence of rogue wave / breaker and MI. However, a sufficiently strong MI will mask the evolution of a rogue wave/breather as the noise grows too rapidly. It is natural to conjecture that a compromise is to have 'mild' values of the MI gain spectrum for a rogue wave/breather to be readily observable. Numerical simulations to this effect will be demonstrated in the next section.

## 5. Computational studies on stability for breathers and rogue waves

In these numerical simulations, a pseudospectral method in the time domain and a fourth-order Runge-Kutta scheme with an adaptive step-size control in the spatial domain are employed. The initial condition for the fiber is given by

$$A(0, t) = (1+noise)\, A_{exact}(0, t),\; B(0, t) = (1+noise)\, B_{exact}(0, t), \tag{18}$$

with *noise* = 0.001(1 − 2·Rand). $A_{\text{exact}}(0, t)$ and $B_{\text{exact}}(0, t)$ are the exact amplitudes of the breather / rogue wave solutions at the input end of the fiber (i.e., $z = 0$). The noise is generated by a random variable 'Rand' in the interval [0, 1] and has an amplitude of 0.1% relative to the input wave.

*5.1 Evolution of the Akhmediev breather*

First, we focus on the Akhmediev breather. In the absence of noise, a breather can be formed smoothly at a proper propagation distance. In the presence of noise, a stable breather can be observed only for parameters that lead to weak MI, as shown in Fig. 4 for $\sigma = 0.25$, $\rho = 1$, $\omega_0 = 1$, and $s = 1.2$, where the maximum gain is 0.2. The appropriate characterization of 'weak' versus 'strong' gain will be made precise after all the numerical results have been presented. When the MI effect is strong, the breather is severely affected by the growth of the background noise (Fig. 5, where the parameters are the same as those for Fig. 4 except for $\sigma = 1$ and the maximum gain is then 0.45).

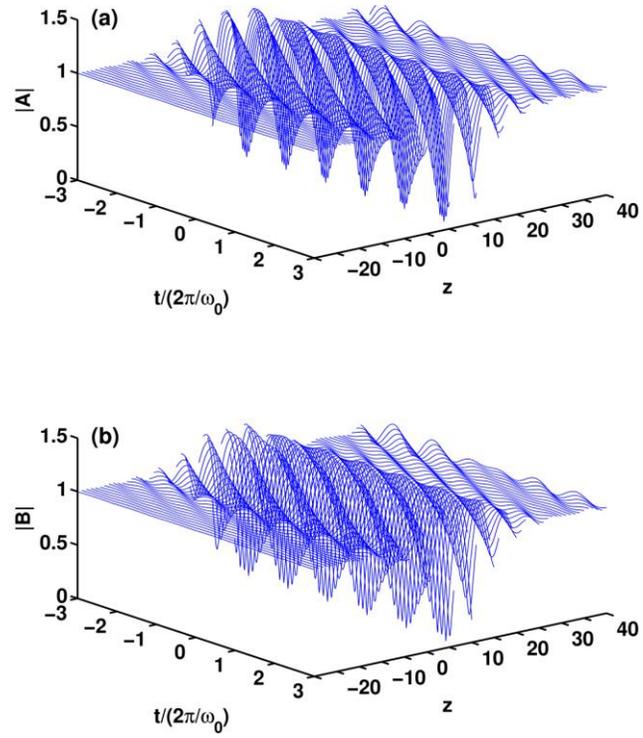

FIG. 4. (Color online) Evolution of the amplitude (a) |A| and (b) |B| with a background noise of 0.1% for the parameters $\rho = 1$, $\omega_0 = 1$, $s = 1.2$, and $\sigma = 0.25$, showing the stable evolution of an Akhmediev breather.

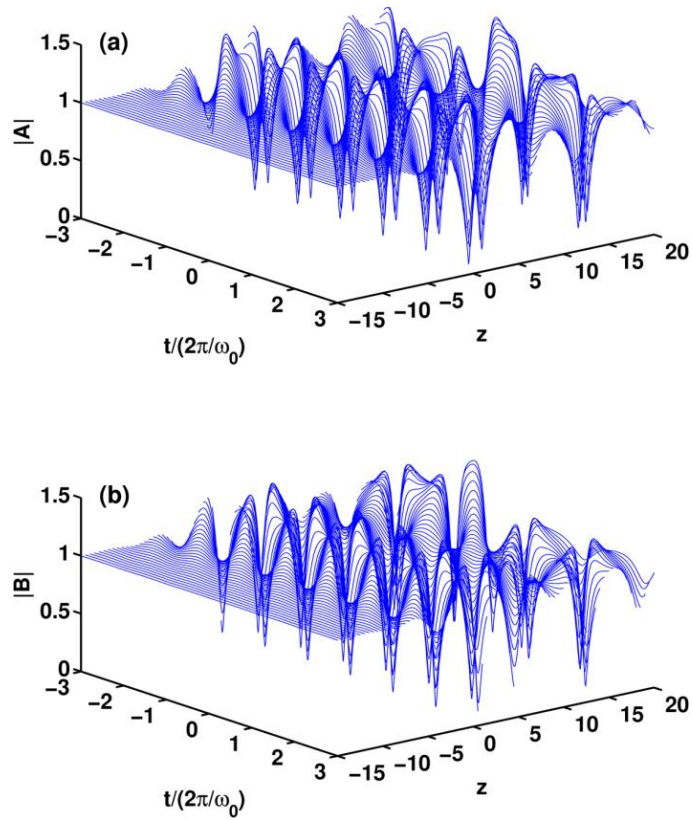

FIG. 5. (Color online) Evolution of the amplitude (a) $|A|$ and (b) $|B|$ with a background noise of 0.1% for the parameters $\rho = 1$, $\omega_0 = 1$, $s = 1.2$, and $\sigma = 1$. The gain displayed by the noise will severely affect or even overwhelm the Akhmediev breather.

For the present scaled version of coupled NLSEs (Eq. (1)), more extensive simulations show that a modulation instability gain in the range of 0.2 to 0.3 will be a threshold for a structural stability of these breathers, above which the pulsating modes will be 'contaminated' by the growth of the noise.

*5.2 Stability of rogue waves*

A similar picture emerges for rogue waves, which is possible only for parameter regimes with MI. In other words, a 'mild' MI will permit a readily observable rogue wave, while a 'strong' MI will 'overwhelm' or 'mask' the development. To illustrate the situation, the evolution with maximum MI gains of 0.10, 0.11, and 0.27 are shown in Figs. 6, 7 and 8 respectively. When the MI effect is weak, a rogue wave can be readily observed (Fig. 6). As the MI gain increases, the rogue wave begins to be affected by MI (Fig. 7). As the MI effect becomes sufficiently strong, the evolution of the rogue wave will be 'masked' by the background instability (Fig. 8). The maximum MI gain which permits a distinct recognition of rogue wave development under the present scaling (Eq. (1)) is found from more extensive simulations to be around 0.1, roughly half the threshold value of the case of the Akhmediev breather.

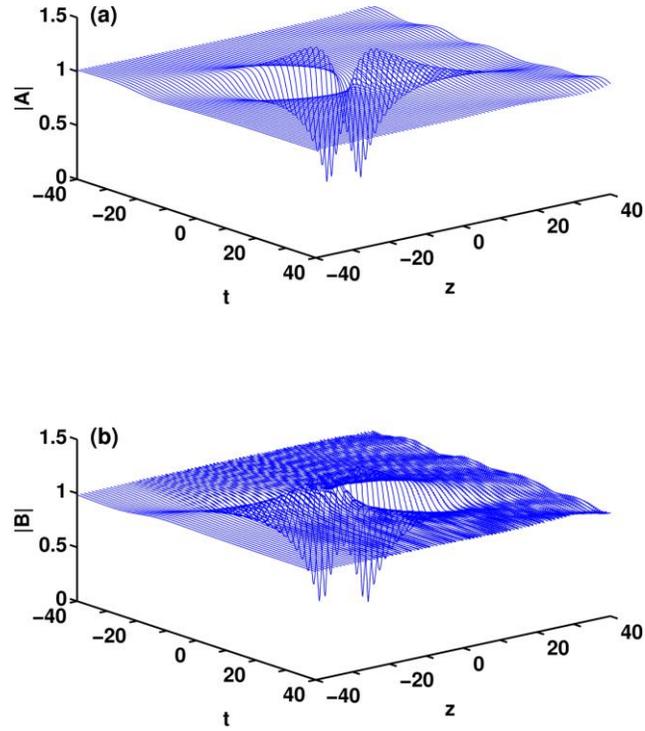

FIG. 6. (Color online) Evolution of the amplitude (a) |A| and (b) |B| with a background noise of 0.1% for $\rho = 1$, $s = 0.5$, and $\sigma = 0.5$, showing the formation of a rogue wave with almost no influence from the background modulation instability.

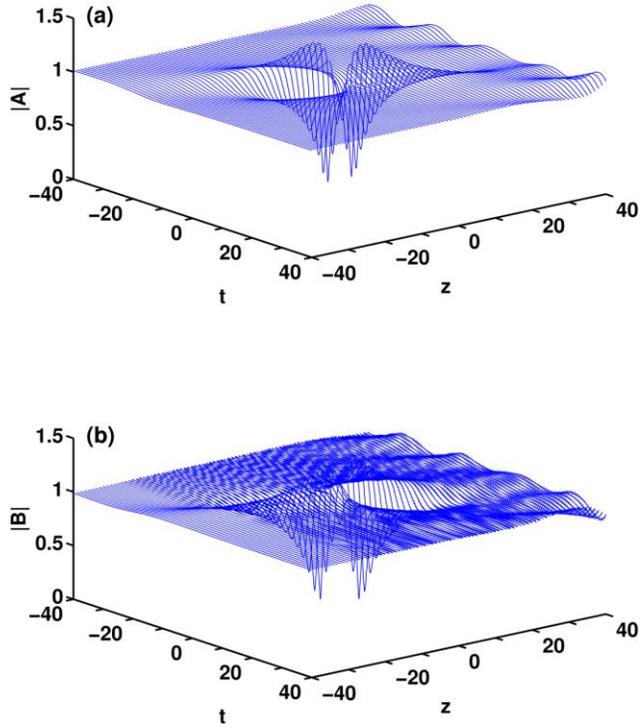

FIG. 7. (Color online) Evolution of the amplitude (a) |A| and (b) |B| with a background noise of 0.1% for $\rho = 1$, $s = 0.5$, and $\sigma = 1$. A few ripples can be seen in the background, but the rogue wave is mostly unaffected.

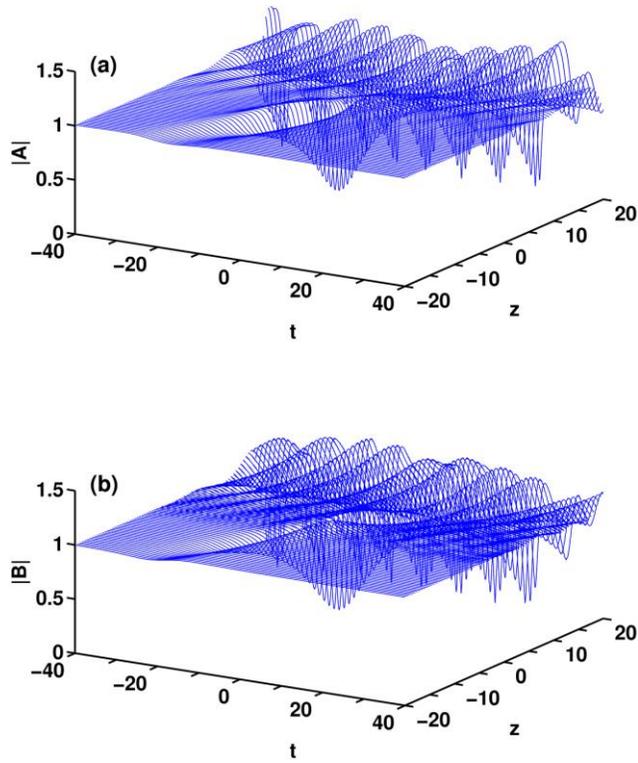

FIG. 8. (Color online) Evolution of the amplitude (a) |A| and (b) |B| with a background noise of 0.1% for $\rho = 1$, $s = 1$, and $\sigma = 0.5$. The rogue wave is 'masked' or 'overwhelmed' in view of the strong modulation instability of the background.

## 6. Discussions and conclusions

Breathers (pulsating modes) have been studied earlier in the literature for coupled nonlinear Schrödinger systems (the Manakov model) in the anomalous dispersion regime [34], and these modes are shown in this paper to exist in the normal dispersion regime as well. The background continuous waves have different frequencies and wavenumbers which can analytically be transformed to a setting of wave packets with group velocity mismatch. These formulations serve as adequate models for the propagation of two orthogonal polarization modes of an optical fiber with randomly varying birefringence. In a sense the present effort generalizes recent works [21, 22] which treats rogue waves for coupled NLSEs in the normal dispersion regime too, as the long wave limits of breathers will yield the rogue wave modes. Theoretically the Hirota bilinear method, instead of the Darboux transformation, is used here, with the merit that the Hirota method has been used in soliton calculations for over forty years, and most 'integrable' equations have bilinear forms.

For the special case of equal background amplitude, it was established [22 and calculations here] that localized modes must have a depression in the center. In addition we identify the existence of a black rogue wave (intensity dropping to zero), and elucidate theoretically the transformation from a configuration of a single 'depression' to two 'depressions'. With two elevations in the vicinity to maintain the conserved quantities, the transformation to a four-petal configuration is elucidated.

The existence criterion for rogue waves correlates exactly with the onset of modulation instabilities. However, an extremely strong background instability may overwhelm or mask the growth phase of the rogue wave. As such a 'mild' range of background should be beneficial to the experimental observation of rogue waves. Computational studies of stabilities conducted confirm this trend. More precisely, for relatively 'weak' MI, both the breathers and the rogue waves can evolve stably in the presence of small perturbations. However, the evolution is distorted or masked if the MI is sufficiently strong.

Quantitatively, in the scaling defined by Eq. (1), the maximum MI gain values that can lead to the stable evolution of Akhmediev breathers and rogue waves should be less than 0.24 and 0.1 respectively. To convert into actual values relevant to an optical fiber, we assume a characteristic length equal to $(T_0)^2/|\beta_2|$ [5] where a typical period $T_0$ of say 1ps is chosen. If the second order group velocity dispersion $\beta_2$ takes on a value of 20ps$^2$/km, these critical thresholds correspond to actual MI gains of 4.8/km and 2/km respectively, which are realistic values in a laboratory setting.

In principle, these studies can be readily extended to other settings of optical rogue waves. Effects of other realistic optical fibers, e.g. multi-mode fibers, can be considered too [35]. A tremendous amount of valuable information is awaiting future efforts of researchers.

**Acknowledgements**

Partial financial support for H.N.C. and K.W.C. has been provided by the Research Grants Council General Research Fund contract HKU 711713E.